\documentstyle[10lomcon,cite,epsf]{article}

\begin{document}
\def\eps{\varepsilon}
\def\eprime{\varepsilon^\prime}
\newcommand{\epr}{\varepsilon^\prime/\varepsilon}
\newcommand{\amp}[1] {\mathrm{A}({#1})}
\newcommand{\repr}{\Re e\,(\epr)}
\newcommand{\pipi}{\pi\pi}
\newcommand{\pio}{\pi^0}
\newcommand{\piopio}{\pi^0\pi^0}
\newcommand{\pippim}{\pi^+\pi^-}
\newcommand{\kethree}{\mathrm{K}_{e3}}
\newcommand{\kmuthree}{\mathrm{K}_{\mu3}}
\newcommand{\ptprime}{p_t{}^\prime}
\newcommand{\asp}{\mathcal{A}}
\def\KL{\rm K_{\rm L}^0}
\def\KS{\rm K_{\rm S}^0}
\def\KZ{\rm K^0}
\def\KZbar{\rm \overline{K}^0}
\def\R{\mathcal{R}}
\def\pio{\pi^{0}}
\newcommand{\CP}{\mbox{$\cal CP$}}
\def\epm{\eta_{+-}}
\def\eoo{\eta_{00}}
\newcommand{\Br}[1] {\mathrm{\Gamma}({#1})}
\newcommand{\E}[1] {\mathrm{\times 10^{#1}}}

\title{The precise determination of $\repr$ }

\author{ M.Cirilli \footnote{e-mail: Manuela.Cirilli@cern.ch}}

\address{University of Rome and INFN, I-00185, Rome, Italy}

\maketitle\abstracts{
\CP\ violation in the neutral kaon system is
known to be dominated by the mixing of $\KZ$ and $\KZbar$.  Direct \CP\
violation in the 2 pion decays of neutral kaons has been a
controversial subject over the last decade.  A strong experimental
effort has been devoted to the precise measurement of the direct \CP\
violation parameter $\repr$.  After 10 years of detector development,
data collection and analysis, the NA48 experiment at CERN and the KTeV
experiment at Fermilab have now established direct \CP\ violation as a
fact.  Both KTeV and NA48 use the same experimental principle,
measuring the double ratio of long lived and short lived neutral kaons
to two charged and two neutral pions.  However, their experimental and
analysis techniques differ in important ways, and I will extensively
discuss the two approaches.  I will also present the latest results on
$\repr$ from both experiments, which were announced just a few months
ago.}

\section{Introduction}
The violation of \CP\ symmetry was first reported in 1964 by J.H.\
Christenson, J.W.\ Cronin, V.\ Fitch and R.\ Turlay, who detected a
clean signal of \CP\ violating $\KL\rightarrow\pippim$
decays~\cite{disco}.  
\CP\ conservation implies that the $\KS$ and $\KL$ particles are pure \CP\
eigenstates and that $\KL$ decays only into \CP$=-1$ and $\KS$ into
\CP$=+1$ final states.  The observed signal of the forbidden
$\KL\rightarrow \pipi$ decays (\CP$=+1$) indicates that \CP\ is not a
conserved symmetry.

\CP\ violation can occur via the mixing of \CP\ eigenstates, called
{\em indirect\/} \CP\ violation, represented by the parameter $\eps$. 
\CP\ violation can also occur in the decay process itself, through the
interference of final states with different isospins.  This is
represented by the parameter $\eprime$ and is called {\em direct\/}
\CP\ violation.  L.\ Wolfenstein in 1964~\cite{wolf} proposed a
super-weak force responsible for $\Delta S=2$ transitions, so that all
observed \CP\ violation phenomena come from mixing and $\eprime =0$. 
In 1973, Kobayashi and Maskawa proposed a matrix representation of the
coupling between fermion families~\cite{kob}.  In the case of three
fermion generations, both direct and indirect \CP\ violation are
naturally accommodated in their model, via an irreducible phase.

The parameters $\eps$ and $\eprime$ are related to the amplitude ratios 
\begin{eqnarray*}
\epm = \frac{ \amp{\KL \rightarrow \pippim} }{ \amp{\KS \rightarrow 
\pippim} } = \eps +  \eprime   
\end{eqnarray*}
and 
\begin{eqnarray*}
\eoo = \frac{ \amp{\KL \rightarrow \piopio} }{ \amp{\KS \rightarrow 
\piopio} } = \eps - 2\eprime
\end{eqnarray*}
which represent the strength of the \CP\ violating amplitude with
respect to the \CP\ conserving one, in each mode.  By the mid-1970s,
experiments had demonstrated that \CP\ violation in the neutral kaon
system is dominated by mixing, with the limit $\repr \le
10^{-2}$\cite{pioneer}.  On the other hand, theoretical work showed
that direct \CP\ violation in the Standard Model could be large enough
to be measurable \cite{firstthe}.  This stimulated experimental effort
with sophisticated detectors to measure $\repr$.  The first evidence
for the existence of a direct component of \CP\ violation was
published in 1988~\cite{evidence}.  In 1993, two experiments published
their final results without a conclusive answer on the existence of
this component.  NA31~\cite{na31} measured $\repr=(23.0\pm6.5)\E{-4}$,
indicating a $3.5\sigma$ effect.  The result of E731~\cite{e731},
$\repr=(7.4\pm5.9)\E{-4}$, was instead compatible with no effect. 

The controversial results from NA31 and E731 called for the
realization of more precise experiments, to measure $\repr$ with a
precision of $\cal O$$(10^{-4}$.  Presently, there are three
experiments in different laboratories working on the precise
measurement of $\repr$: two of these, namely NA48~\cite{na48prop} at
CERN and KTeV~\cite{ktevreport} at Fermilab, represent the
``evolution'' of NA31 and E731 respectively; the third one is
KLOE~\cite{kloetp} at the Laboratori Nazionali di Frascati and its
conceptual design is radically different from the other experiments. 

This paper is devoted to a comparative presentation of KTeV and NA48, 
since these experiments have already published results. Detector 
designs and analysis techniques will be discussed, together with the 
results announced just recently from these two collaborations 
\cite{na48:eprime2001}\cite{ktev:eprime2001}.

\section{The experimental method}\label{sec:method}
Experimentally, it is convenient to measure the double ratio $\R$,
which is related to the ratio $\repr$:
\begin{equation}
R = \frac{ \Br{\KL \rightarrow \piopio} }{ \Br{KS \rightarrow \piopio}
} / \frac{ \Br{\KL \rightarrow \pippim} }{ \Br{\KS \rightarrow
\pippim} } \approx 1 - 6 \times \repr
\label{doubleratio}
\end{equation}
The double ratio $\R$ is experimentally measured by \textit{counting}
the number of decays detected in each of the four modes in
equation~\ref{doubleratio}.  The statistical error is dominated by the
events collected in the most suppressed decay, namely
$\KL\rightarrow\piopio$ ($BR\sim 0.09$\%).  The value $\R$$_{true}$ is then
deduced correcting the measured value $\R$$_{meas}$ for the kaon beam
fluxes, detector acceptances, trigger efficiencies, backgrounds
evaluations, etc., i.e. for all the possible biases in the counting
process.  It is now evident that the difficulty of $\repr$
measurements lies in the necessity to disentangle the \CP\ violating $\KL$
modes from the dominant environment of \CP\ conserving 3-body decays
of both $\KL$ and $\KS$.

\subsection{Advantages of the double ratio technique}\label{subsec:adv}

The main advantage of the double ratio measurement, when performed
under the adequate data taking conditions, is that the corrections to
$\R$$_{meas}$ can cancel out at first order.Let us consider the beam fluxes
and trigger/reconstruction efficiencies corrections as an example:
\begin{itemize}
    \item \textbf{Beam fluxes:} the knowledge of the kaon flux in the
    $\KS$ and $\KL$ beams is a priori needed for normalization
    purposes.  However, if the charged and neutral decay modes of
    either the $\KS$ or the $\KL$ are simultaneously collected, then
    the ratio of $\pippim$ and $\piopio$ events in each beam is
    independent from the absolute flux.  Hence, under these
    conditions, beam fluxes cancel out in the double ratio at first
    order.

    \item \textbf{Efficiencies:} the trigger scheme is conceived to
    minimise any loss of good events.  However, a small correction
    usually has to be applied to $\R$$_{meas}$ to account for trigger
    inefficiencies.  A first order cancellation of this correction can
    be achieved for the charged/neutral trigger efficiency if both
    $\KL$ and $\KS$ decays into the charged/neutral final state are
    simultaneously collected.  The same principle also holds for any
    instability of a given detector, which could affect the
    reconstruction efficiency of the charged or neutral modes.
\end{itemize}
The best strategy to exploit the cancellation of eventual biases is to
collect all the four modes simultaneously.  This allows to evaluate
only second order effects to get the true value $\R$$_{true}$ from
$\R$$_{meas}$.  Even in this ideal situation, there will still be some
leftover corrections that do not cancel out.  This is the case for the
physical background, which comes only from $\KL$ decays and is clearly
final-state-dependent.  Also, acceptance corrections do not a priori
cancel out in the four modes: this is related to the huge lifetime
difference between $\KL$ and $\KS$, which causes very different
longitudinal decay vertex distributions for the two beams, and to the
different topologies of $\piopio \rightarrow 4\gamma$ and $\pippim$
events. Both the physical background and the acceptance correction 
must be carefully studied, and different solutions can be envisaged 
to handle them.

All the above considerations have been thoroughly taken into account
while conceiving KTeV and NA48.  The design of the experiments and the
analysis methods focus on making the inevitable systematic biases in
the event counting symmetric between at least two of the four
components of the double ratio.  In this way, most of the important
systematic effects cancel to first order, and only the differences
between two components need to be considered in detail in the
analysis.  This allows the systematic uncertainties to be kept
sufficiently low.

\section{KTeV and NA48: overview}\label{sec:overview}
Both KTeV and NA48 are fixed target experiments designed to
simultaneously collect all the four decay modes in \ref{doubleratio}. 
Measuring $\repr$ to a precision of $\sim 10^{-4}$ requires several
millions of  $\KL$ and $\KS \rightarrow \pipi$ decays: this implies taking
data with high-intensity kaon beams and running for several years to
achieve the desired statistics.  A number of challenges had to be
faced during the design phase.  Stable detectors were needed to
sustain the long data taking periods.  The trigger electronics had to
be fast enough to cope with the high flux of particles in the decay
region and a powerful data acquisition was needed to handle the high
trigger rates.  The overwhelming $3\pio$ background set the
requirement of an extremely precise electromagnetic calorimeter.  The
whole detector had to be radiation-hard, to cope with the beam
intensities.  Long R\&Ds were necessary to meet these very demanding
requirements, and this effort has also been profitable in view of future
experiments in high-intensity environments (\textit{e.g.}, the Tevatron and LHC).

KTeV collected $\sim 7$M events in the most suppressed channel $\KL
\rightarrow \piopio$ during the 96, 97 and 99 runs.  A first $\repr$
measurement \cite{ktev:eprime99} was announced in February, 1999,
based on $10$\% of the total sample.  In June, 2001, KTeV presented a
new result on the 97 data sample, together with an update of the
already published result: the combined measurement
\cite{ktev:eprime2001} is obtained from $\sim 50$\% of the available
data.  NA48 took data in 97, 98 and 99, collecting almost $4$M events
in the neutral $\KL$ decay mode.  The first $\repr$ measurement
\cite{na48:eprime99} was reported in June, 1999, and was based on the
statistics collected during the 97 run; a preliminary result on the 98
sample was presented in February, 2000.  The final result on the 98
and 99 data~\cite{na48:eprime2001} was announced in May, 2001.  A
slight increase in statistics for NA48 is expected from the data
collected in the 2001 run.

\subsection{KTeV: Detector and beam lines}\label{subsec:ktev}
\begin{figure}[t!]
\begin{center}
\epsfxsize=9.5cm
\epsffile{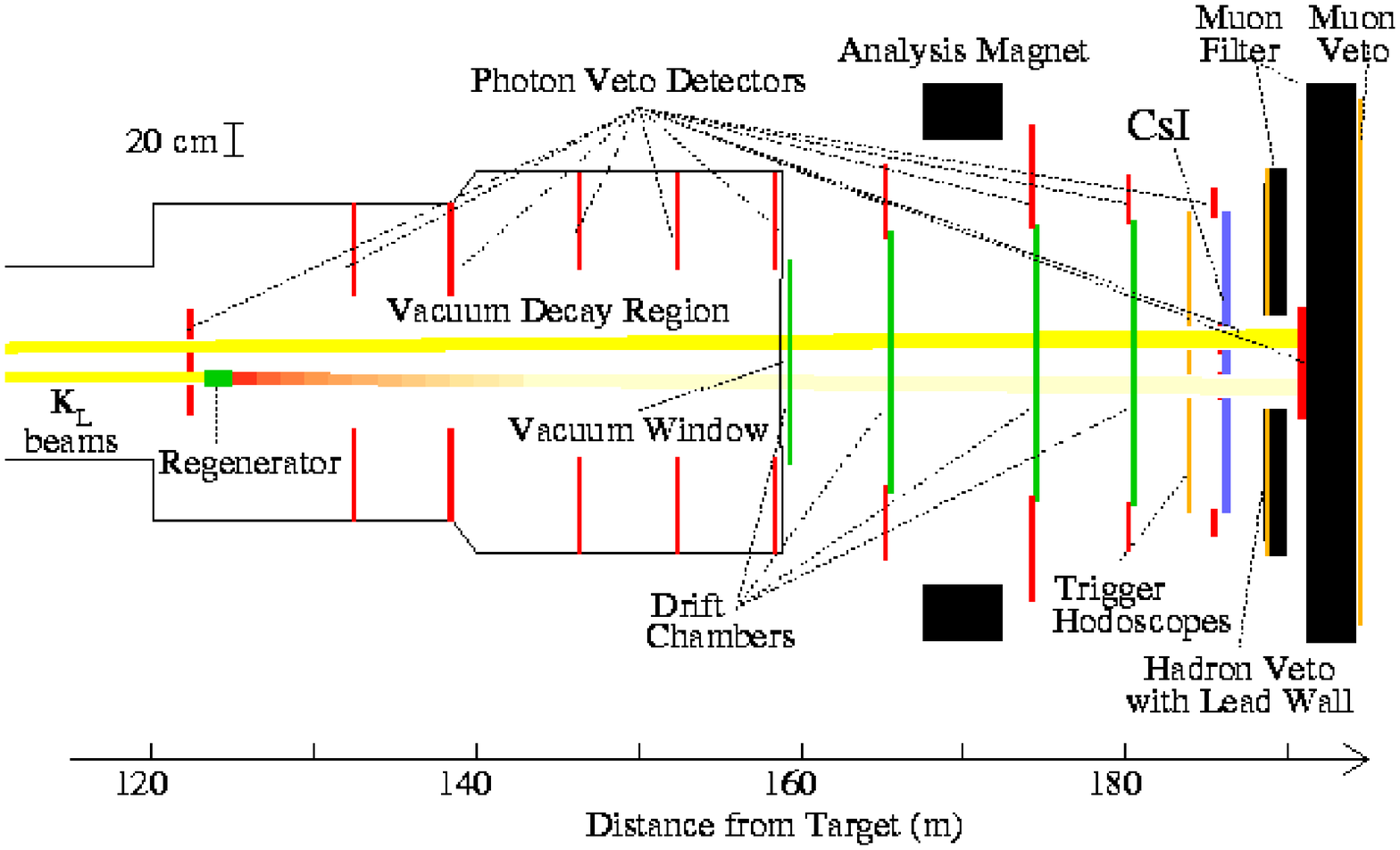}
\caption{Top view of KTeV detector and beam lines.}
\label{fig:ktevsetup}
\end{center}
\end{figure}

A schematic view of KTeV detectors and beams is shown in
figure~\ref{fig:ktevsetup}.  KTeV exploits the $800$~GeV proton beam
delivered by the Tevatron: two nearly parallel kaon beams are produced
by the protons hitting a $50$~cm long beryllium target at $4.8$~mrad
angle.  The beams are cleaned up and let fly for roughly $120$~m, so
that only the $\KL$ component survives.  The beams¹ direction defines
the longitudinal $z$ axis.  The decay region begins at the end of the
last collimator; here the two beams are $10$~cm apart, and one of them
hits a $1.8$~m long regenerator made of plastic scintillators.  The
regenerator beam is a coherent superposition $\KL +\rho \KS$ of long-
and short-lived kaons.  The regenerated fraction $\rho$ is
proportional to the amount of matter traversed by the previously pure
$\KL$ beam, and its value $0.03$ is sufficient to ensure that the
regenerator $2\pi$ decays are dominated by $\KS \rightarrow \pipi$. 
The regenerator technique ensures that the $\KS$ are produced with an
energy spectrum similar to that of $\KL$.  The decay region extends up
to $159$~m from the primary target.

%%%%%%%%%%%%%%%%%%%%%%%%%%%%%%%%%%%%%%%%%%%%%%%%%%%%%%%%%
%%%%%%%%%%%%%%%%%%%%%%%%%%%%%%%%%%%%%%%%%%%%%%%%%%%%%%%%%

A distinctive feature of KTeV is the fact that the two beams are
parallel and hit the detectors at separate points (left and right). 
This allows to easily identify $\KL$ and $\KS$ decays reconstructing
the transverse decay vertex position and comparing it with the known
regenerator position.  The regenerator is fully instrumented, and
switches beam line once per minute, in order to reduce the effects of
possible left-right asymmetries of the detectors.

Charged kaon decays are detected by a spectrometer consisting of a 
central magnet with a $411$~MeV/$c$ kick in the horizontal plane and 
of four drift chambers with wires along the $x$ and $y$ directions. 
The spectrometer has a position resolution of $100 \mu$m and a 
momentum resolution $\sigma_{p}/p = 0.17\% \oplus [0.008 \times 
p]\%$, where $p$ is in GeV/$c$ units.  

\begin{figure}[hbtp]
\begin{center}
\epsfxsize=4.cm
\epsffile{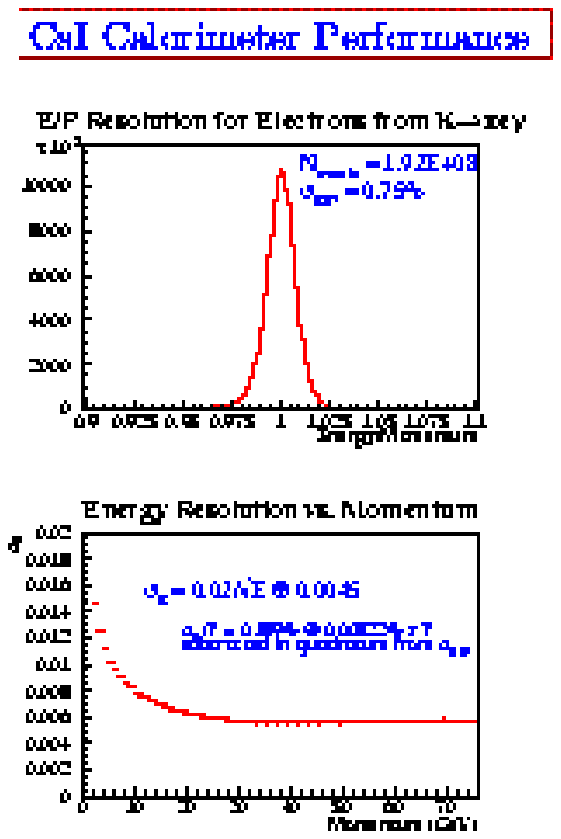}
\epsfxsize=5.5cm
\epsffile{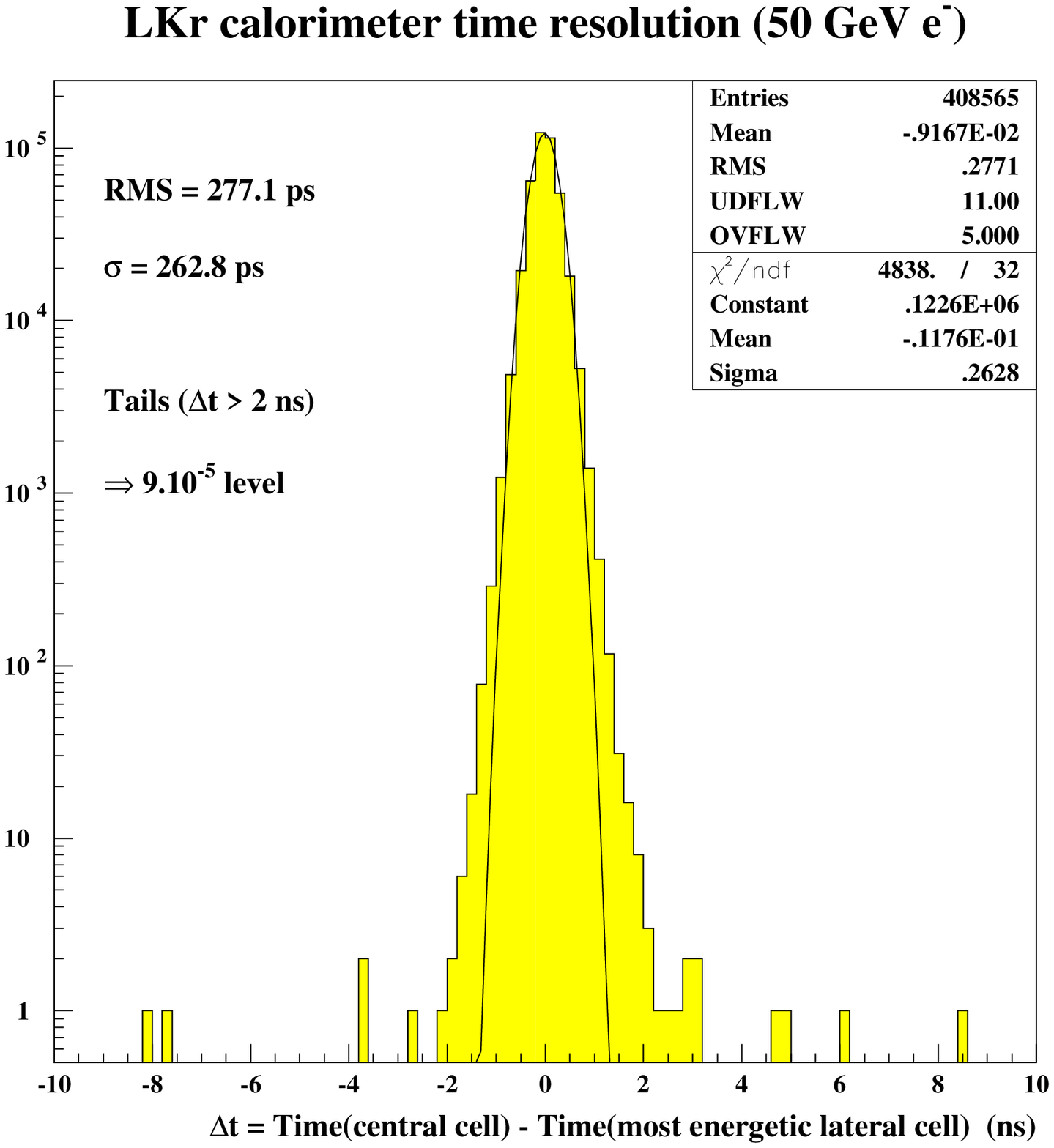}
\caption{Performances of KTeV Csi crystal calorimeter (left) and of the
  NA48 LKr ionization calorimeter (right)}
\label{calor}
\end{center}
\end{figure}

Neutral decays are detected by a crystal calorimeter~\cite{ktev:calo}
consisting of $3\,100$ pure CsI blocks.  The crystals cover $27
\mathrm{X}_{0}$ in length ($50$~cm) and have a transverse section of
$2.5\times 2.5 \mathrm{cm}^{2}$ in the central region, where the
density of photons is higher; in the outer region of the calorimeter,
the granularity is of $5\times 5 \mathrm{cm}^{2}$.  The main advantage
of this calorimeter lies in its excellent stochastic term in the
energy resolution, which allows to reach an overall resolution of
$0.7$\% for a $15$ GeV photon (as shown in figure~\ref{calor} left).  The
longitudinal light collection is equalised within $5$\% by means of a
meticulous crystal wrapping.  In addition, the stability of the
response for each crystal is continuously checked using a
$\mathrm{Cs}^{137}$ source for calibration.  The overall  
energy response is linear within $0.4$\%.

%%%%%%%%%%%%%%%%%%%%%%%%%%%%%%%%%%%%%%%%%%%%%%%%%%%%%%%%%
%%%%%%%%%%%%%%%%%%%%%%%%%%%%%%%%%%%%%%%%%%%%%%%%%%%%%%%%%

The main apparatus is surrounded by circular vetoes to detect escaping 
photons, and a muon veto placed at the end of the line is used to 
identify $\kmuthree$ decays.

\subsection{NA48: Detector and beam lines}\label{subsec:na48}
\begin{figure}[hbtp]
\begin{center}
\epsfxsize=10.cm
\epsffile{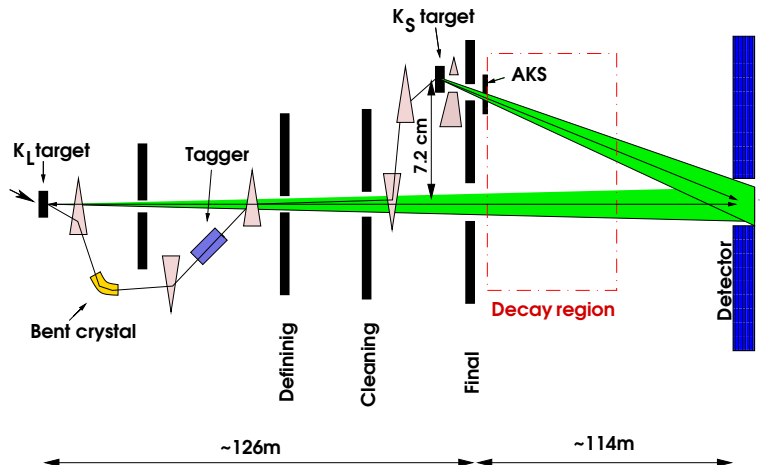}
\caption{Schematic view of NA48 beam lines (not to scale).}
\label{na48beams}
\end{center}
\end{figure}

A schematic view of the NA48 beam lines is given in
figure~\ref{na48beams}.  The primary $450$ GeV proton beam is
delivered from the SPS and impinges on a $40$~cm beryllium target with
an incidence angle of $2.4$~mrad relative to the $\KL$ beam axis.  The
charged component of the outgoing particles is swept away by bending
magnets, while the neutral beam component passes through three stages
of collimation.  The fiducial region starts at the exit of the
``final'' collimator, $126$~m downstream of the target.  At this
point, the neutral beam is dominated by long-lived kaons.  The
non-interacting protons from the $\KL$ target are directed onto a
mechanically bent mono-crystal of silicon.  A small
fraction ($10^{-5}$) of protons satisfies the conditions for
channelling and is deflected following the crystalline planes.  Use of
the crystal allows a deflection of $9.6$~mrad to be obtained in only
$6$~cm length, corresponding to a bending power of $14.4$~Tm.  The
transmitted protons pass through the tagging station (or
\textit{tagger}), which precisely registers their time of passage. 
They are then deflected back onto the $\KL$ beam axis, transported
through a series of quadrupoles and finally directed to the $\KS$
target (same size as $\KL$) located 72~mm above the $\KL$ beam axis. 
A combination of collimator and sweeping magnet defines a neutral beam
at 4.2~mrad to the incoming protons.  The decay spectrum of kaons at
the exit of the collimator is similar to that in the $\KL$ beam, with
an average energy of 110~GeV. The fiducial region begins 6~m
downstream of the $\KS$ target, such that decays are dominated by
short lived particles.  At this point, the $\KS$ and $\KL$ beams
emerge from the aperture of the final collimators into the common
decay region.  The whole $\KS$ target and collimator system is aligned
along an axis pointing to the centre of the detector 120~m away, such
that the two beams intersect at this point with an angle of 0.6~mrad.

%%%%%%%%%%%%%%%%%%%%%%%%%%%%%%%%%%%%%%%%%%%%%%%%%%%%%%%%%
%%%%%%%%%%%%%%%%%%%%%%%%%%%%%%%%%%%%%%%%%%%%%%%%%%%%%%%%%

Since the two beams are not separated at the detector position, as it
is in KTeV, the identification of $\KL$ and $\KS$ decays must be
accomplished in a different way.  This is done using the tagging
station, which consists of two scintillator ladders, crossing the beam
horizontally and vertically.  The coincidence between the proton time
and the event time in the detectors assigns the decay to the $\KS$
beam.  Two close pulses can be resolved down to 4--5~ns.

The reconstruction of charged decays is performed by a magnetic
spectrometer, with a central magnet giving a $250 \mathrm{MeV}/c$
transverse kick and four chambers with plane wires oriented along 
four different directions $x$, $y$, $u$ and $v$. The redundancy of the 
planes allows to resolve the possible track ambiguities.

NA48 has chosen a liquid Krypton ionization calorimeter with a depth
of $27 \mathrm{X}_{0}$, corresponding to $125$~cm.  The read-out is
performed by Cu-Be-Co ribbons defining $\sim 13\,000$ cells, in a
structure of longitudinal projective towers pointing to the centre of
the decay region.  The cross section of a cell is about 2~cm $\times$
2~cm, and the electrodes are guided longitudinally through precisely
machined holes in five spacer plates.  The planes also apply a
$\pm$48~mrad zig-zag to the electrodes, in order to maintain the
mechanical stability and to decrease the sensitivity of the energy
resolution to the impact position.  Good energy response is further
guaranteed by the initial current readout technique which also
provides a high rate capability.  The overall energy resolution is
$1.5$\% at $10$~GeV. The energy response is linear to about 0.1\% in
the range 5--100~GeV. A precise time measurement is mandatory for the
NA48 calorimeter, since it must be used together with the proton time
from the tagger to distinguish $\KS$ from $\KL$.  The neutral event
time is reconstructed with a precision of $\sim 220$~ps.; tails coming
from misreconstructed times are below the level of 10$^{-4}$ (see 
figure~\ref{calor} right).

A muon veto system is used to reject muons from $\kmuthree$ decays.

\section{KTeV and NA48: analysis techniques}\label{sec:anal}
Once the events are collected, all corrections that do not cancel in
the double ratio must be applied.  The long list of residual effects
that must be studied includes physical backgrounds, $\KS -\KL$
misidentifications, trigger efficiencies, Monte Carlo correction,
geometrical acceptances, detector biases (calorimeter energy scale,
drift chamber alignment, etc.), accidental effects. In the following, 
I will focus on just a few of these effects, highlighting the 
important differences between KTeV and NA48 approach.

\subsection{Selection of the $\pippim$ sample}\label{subsec:char}
Both experiments use the magnetic spectrometer to reconstruct the kaon
mass, vertex and momentum.  The resolution on the kaon mass in the
charged mode is $1.5$~MeV in KTeV and $2.5$~MeV in NA48.  The better
KTeV resolution is due to the higher transverse kick of their magnet
and to the choice of having only $x$ and $y$ planes in the drift
chambers.  This choice implies lighter chambers with respect to NA48,
and so a lower multiple scattering term in the resolution.  However,
having only two views reduces the capability of resolving ambiguities
in the track reconstruction: for this reason, KTeV needs additional
information from the calorimeter to perform a reliable reconstruction
of $\pippim$ events. 

\begin{table}[htbp!]
    \centering
    \begin{tabular}{|c|c|c|c|}
        \hline
        \multicolumn{4}{|c|}  {\textsf{Background to $\KZ \rightarrow 
        \pippim$}}  \\
        \hline
        \textsf{Background source} & \textsf{KTeV (vac)} & \textsf{KTeV 
        (reg)} & \textsf{NA48 $\KL$}  \\
        \hline
        $\kmuthree + \kethree$ & $0.9\times 10^{-3}$ & $0.03\times 
        10^{-3}$ & $1.69\times 10^{-3}$  \\
        \hline
        \textsf{Collimator scatt.} & $0.10\times 10^{-3}$ & $0.10\times 10^{-3}$ & -  \\
        \textsf{Regenerator scatt.}  & - & $0.73\times 10^{-3}$ & -  \\
        \hline
    \end{tabular}
    \caption{Summary of background fractions in the charged mode.}
    
    \label{tbl:bkgchar}
\end{table}

$\kmuthree$ decays are rejected using the identified muon in the
dedicated vetoes, while electrons from $\kethree$ events are
identified comparing the track momentum in the spectrometer with the
corresponding energy in the calorimeter.  Additional cuts are imposed
on the reconstructed mass and transverse momentum.  The leftover
background contributions are then evaluated studying high-statistics
samples of the identified 3-body decays.  The background fractions in
the two experiments are summarized in table~\ref{tbl:bkgchar},
including the components due to scattering in the collimator and (only
for KTeV) regenerator.

\subsection{Selection of the $\piopio$ sample}\label{subsec:neu}
Both experiments base the reconstruction of neutral events on the 
information from the electromagnetic calorimeter. In addition to the 
energies and positions of the four photons, a mass constraint must be 
imposed in order to reconstruct the decay vertex position. KTeV 
method imposes the $\pio$ mass to all photon pairs combinations, 
computing the vertex position for each pairing; only the two closest 
solutions are kept, and they are combined to produce the most probable 
value for the kaon decay vertex. NA48 method imposes the kaon 
invariant mass on the $4\gamma$ event, thus constraining the decay 
vertex position. The two $pio$ are reconstructed choosing the best of 
all the possible pairings between the photons.

\begin{table}[htbp!]
    \centering
    \begin{tabular}{|c|c|c|c|}
        \hline
        \multicolumn{4}{|c|}  {\textsf{Background to $\KZ \rightarrow 
        \piopio$}}  \\
        \hline
        \textsf{Background source} & \textsf{KTeV (vac)} & \textsf{KTeV 
        (reg)} & \textsf{NA48 $\KL$}  \\
        \hline
        $\KL \rightarrow 3\pio$ & $1.1\times 10^{-3}$ & $0.3\times 
        10^{-3}$ & $0.59\times 10^{-3}$  \\
        \hline
        \textsf{Collimator scatt.} & $1.2\times 10^{-3}$ & $0.9\times
        10^{-3}$ & $0.96\times 10^{-3}$ \\
        \textsf{Regenerator scatt.} & $2.5\times 10^{-3}$ &
        $11.3\times 10^{-3}$ & - \\
        \textsf{Regenerator had.  int} & - & $0.1\times 10^{-3}$ & -
        \\
        \hline
    \end{tabular}
    \caption{Summary of background fractions in the neutral mode.}
    
    \label{tbl:bkgneu}
\end{table}

Both experiment define a $\chi^{2}$ variable that states the
compatibility of each event with the $\KZ \rightarrow \piopio$
hypothesis.  Background events from $\KL \rightarrow 3\pio$ decays
with lost or merged photons have a high value of the $\chi^{2}$ and
are rejected.  The amount of remaining background is evaluated from a
high-statistic sample of $3\pio$ events.  Background fractions for
KTeV and NA48 are summarized in table~\ref{tbl:bkgneu}.

\subsection{$\KS$ and $\KL$ identification}\label{subsec:tagging}
\begin{figure}[hbtp]
\begin{center}
\epsfxsize=4.6cm
\epsffile{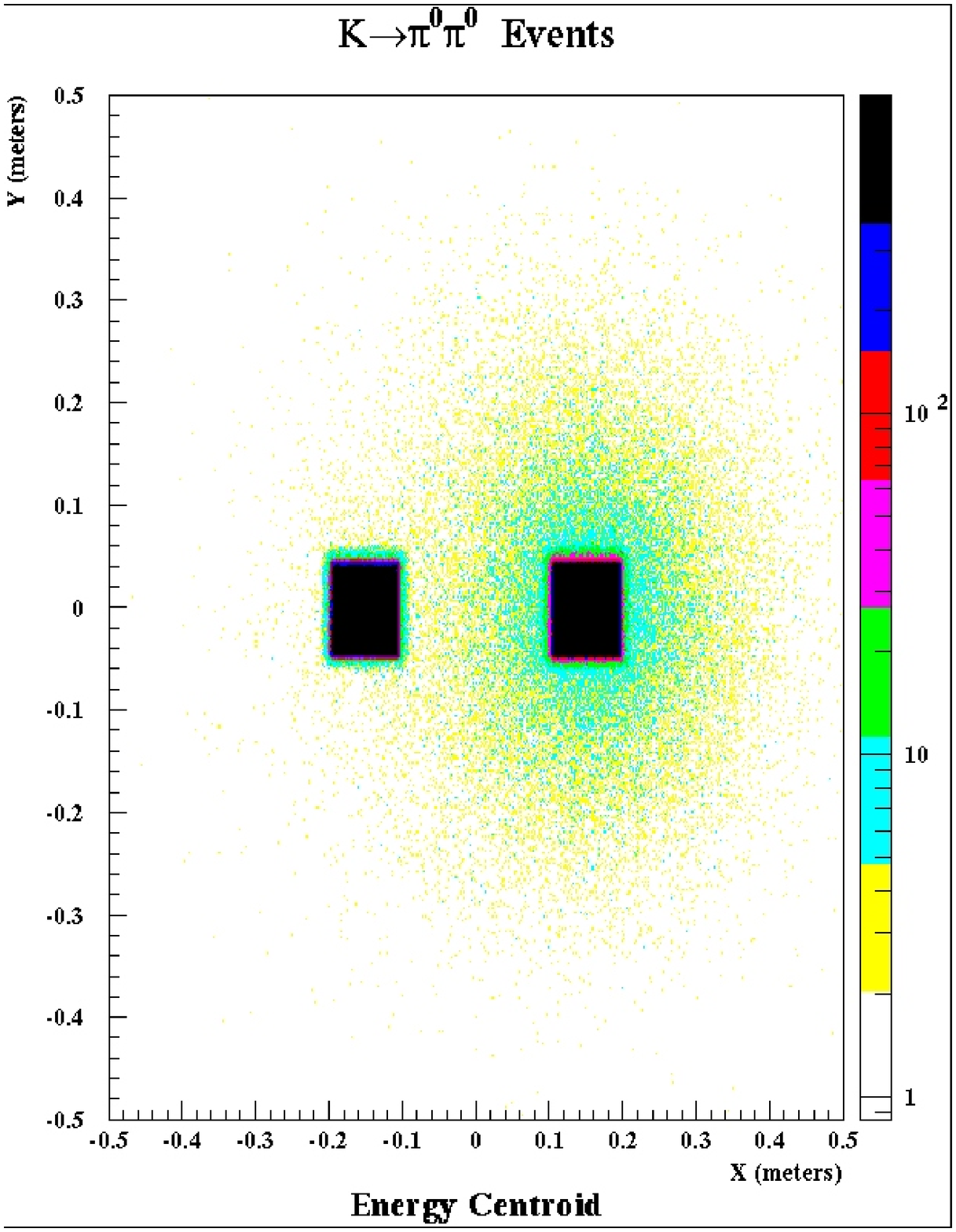}
\epsfxsize=5.cm
\epsffile{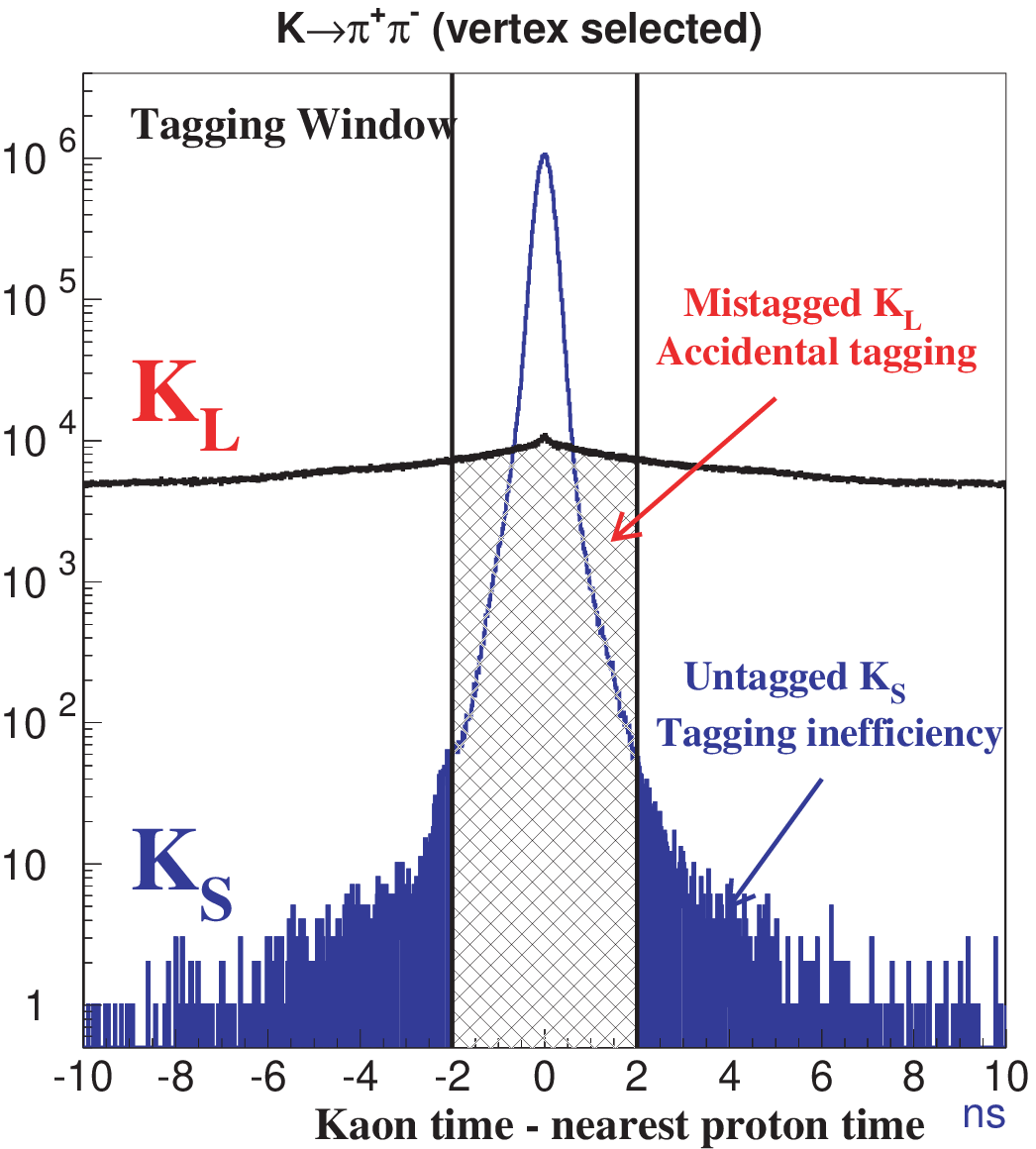}
\caption{Procedure for $\KS$ and $\KL$ identification in KTeV (left) and
  NA48 (right).}
\label{tagging}
\end{center}
\end{figure}

As already described in sections~\ref{subsec:ktev}
and~\ref{subsec:na48}, the two experiments use different techniques to
distinguish $\KS$ from $\KL$.  KTeV takes advantage of having two
parallel beams: the $10$~cm separation allows to disentangle $\KS$ and
$\KL$ by looking at the reconstructed decay vertex position in the
transverse plane, in the case of charged events.  In the case of
neutral events, the energy centroid of the four photons is used, as
shown in figure~\ref{tagging} left.  The halo surrounding one of the two beams
is due to events scattered in the regenerator before decaying: this
effect is accurately studied in the $p_{\mathrm{T}}^{2}$ distribution
of charged events, and is then introduce into a detailed simulation 
to evaluate the contribution in the neutral case.

%%%%%%%%%%%%%%%%%%%%%%%%%%%%%%%%%%%%%%%%%%%%%%%%%%%%%%%%%
%%%%%%%%%%%%%%%%%%%%%%%%%%%%%%%%%%%%%%%%%%%%%%%%%%%%%%%%%

NA48 uses the tagging method: a decay is identified as $\KS$ if its
event time is within a $\pm2$~ns coincidence with a proton time
measured by the tagger.  The principle can be clearly illustrated for
charged events, which can be identified as $\KS$ or $KL$ also on the
basis of the vertical separation between the two beams at the drift
chamber position.  This is shown in figure~\ref{tagging} right, where the
difference between the event time and the time of the nearest proton
in the tagger is plotted.  It is evident that the inefficiencies in
identifying true $\KS$ decays is very small ($10^{-4}$ level).  On the
other hand, there is a sizeable mistagging probability in the case of
$\KL$: the high rate of events in the $\KL$ beam causes accidental
coincidences, and it turns out that $\sim 10$\% of the true $\KL$
events are misidentified as $\KS$.  The final data samples are
corrected for both inefficiency and accidental tagging.

\subsection{From event counting to $\mathcal{R}$}\label{subsec:rcalc}
Having identified $\KS$ and $\KL$, as well as charged and neutral
decays, both experiments end up with four samples of events.  In the
case of KTeV, the samples correspond to the Vacuum beam ($\KL$ decays)
and to Regenerator beam (mostly $\KS$) into $\pippim$ and $\piopio$
final modes. The striking difference in the decay vertex 
distributions for $\KS$ and $\KL$ translates in a large acceptance 
correction, which therefore must be precisely known. The correction 
is implemented using a highly detailed Monte Carlo simulation which 
includes all known effects, as trigger and detector efficiencies, 
regeneration, $\KS -\KL$ interference, detector apertures, etc. The 
value of $\repr$ is then obtained fitting the data in $10$~GeV bins 
in the kaon energy, in order to minimise residual differences in the 
energy spectra.

NA48 has two charged/neutral samples of tagged events (essentially
$\KS$) and two other charged/neutral samples of untagged events ($\KL$
with a $\sim 10$\% contamination of $\KS$).  All samples are corrected
for mistagging and trigger inefficiencies.  The final result is
computed by dividing the data into 20 bins of kaon energy from 70 to
170 GeV, and calculating the double ratio for each bin.  To cancel the
contribution from the different lifetimes to the acceptance, $\KL$
events are weighted with the $\KS$ lifetime as a function of the
reconstructed proper decay time.  After weighting, the $\KL$ and $\KS$
decay distributions become nearly identical and the size of the
acceptance correction is drastically reduced.  The weighting technique
avoids the need of an extremely sophisticated simulation, although it
results in a $\sim 35$\% increase of the statistical error on
$\mathrm{R}$. All corrections are applied to each bin separately, 
and the results are averaged using an unbiased estimator.
%%%%%%%%%%%%%%%%%%%%%%%%%%%%%%%%%%%%%%%%%%%%%%%%%%%%%%%%%
\begin{figure}[hbtp]
\begin{center}
\epsfxsize=5.cm
\epsffile{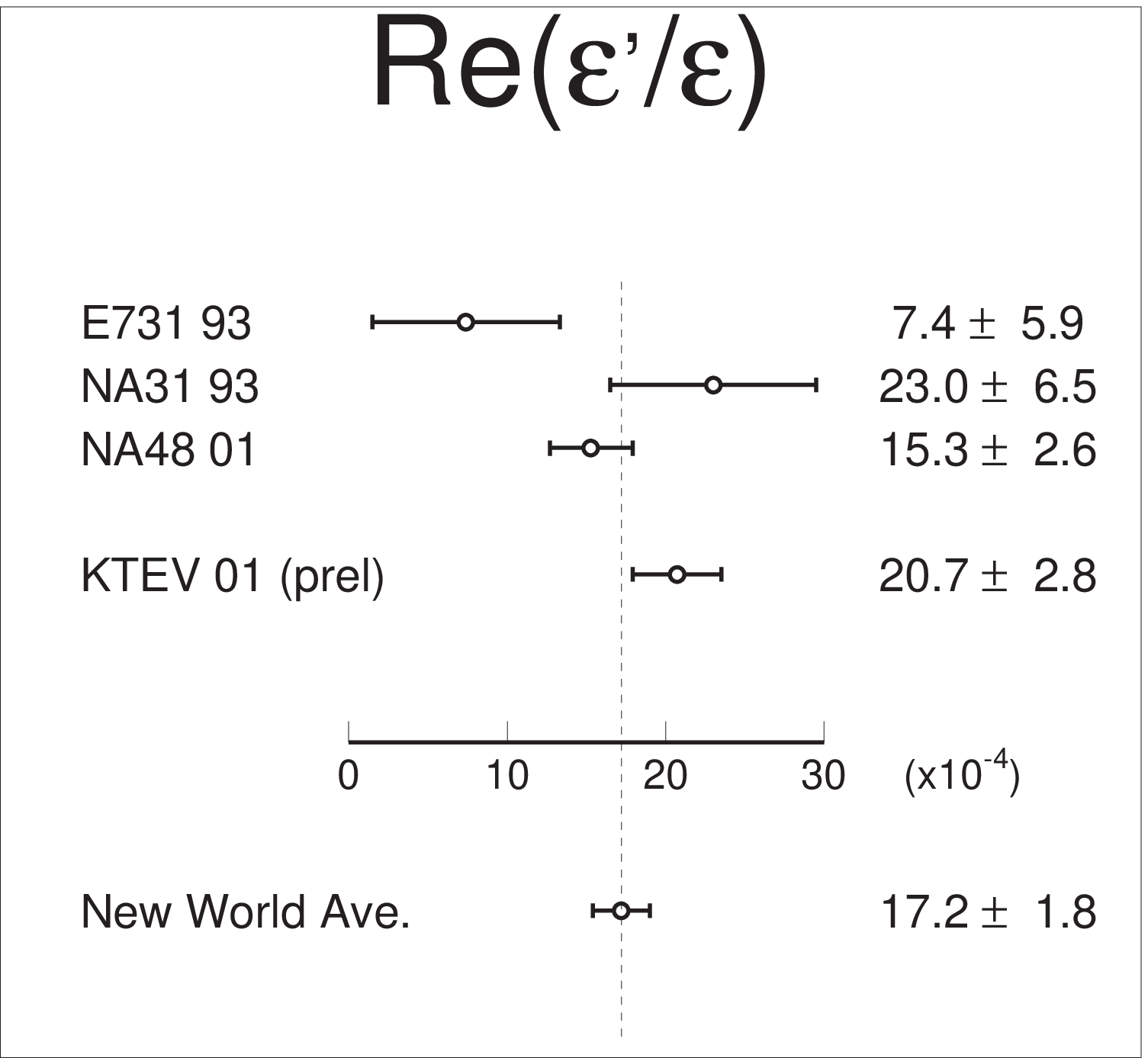}
\caption{Summary of results on $\repr$.}
\label{result}
\end{center}
\end{figure}
%%%%%%%%%%%%%%%%%%%%%%%%%%%%%%%%%%%%%%%%%%%%%%%%%%%%%%%%%

\section{Results from KTeV and NA48}\label{sec:results}
The latest results from KTeV \cite{ktev:eprime2001} and NA48
\cite{na48:eprime2001} are summarized in figure~\ref{result}, together
with the final results from NA31 and E731.  The new world average
value of $\repr$ is $(17.2\pm 1.8)\times 10^{-4}$.  This result
confirms the existence of direct \CP\ violation in the neutral kaon
system.  Whether the measured size of $\repr$ is compatible with
Standard Model expectations or is a hint that new physics is at work,
this is still matter of debate: theoretical calculations suffers from 
big uncertainties in the determination of the hadronix mass matrix 
elements, thus their predictive power on $\epr$ is rather poor. 

Establishing beyond doubt the existence of the direct \CP\ violation
mechanism has been a long experimental adventure.  Both KTeV and NA48
have still other data samples to analyse, and hopefully KLOE will
provide also its measurement of $\repr$ with a different method. 
We recently witnessed the first observation of \CP\ violation in
a system other than the neutral kaon system, namely in ${\rm B^0}
-{\rm \overline{B}^0}$ oscillations. \CP\ violation studies are 
also being performed in the sector of $\mathrm{B}$ and $\mathrm{K}$ rare decays. 
Considering all these constraints together, there is reasonable hope 
that a deeper understanding of the \CP\ violation mechanism will be 
achieved in the forthcoming years.
 
\section*{Acknowledgements}
I would like to warmly thank the organisers of the $10^{th}$ Lomonosov
Conference on Elementary Particle Physics for the interesting meeting.

\end{document}